\documentclass[twocolumn, 12pt]{IEEEtran}
\usepackage{hyperref}
\usepackage[utf8]{inputenc}
\usepackage{amsmath}
\usepackage{amssymb}
\usepackage{graphicx}
\usepackage{multicol}
\graphicspath{{figures/}}
\usepackage{subcaption}
\usepackage{tikz}
\usepackage{siunitx}
\usepackage{booktabs}
\usepackage{multirow}
\usepackage{verbatim} 
\usepackage{pgfplots}
\pgfplotsset{compat=1.15}
\usepackage{listings}

\title{Photo-Voltaic Panel Power Production Estimation with an Artificial Neural Network using Environmental and Electrical Measurements}
\author{Antony~Morales-Cervantes, Oscar~Lobato-Nostroza, Gerardo~Marx~Chávez-Campos, Yvo~Marcelo~Chiaradia-Masselli, and Rafael~Lara-Hernández
\thanks{The submitted paper comes from the research project 6127.17-P funded by the "Tecnológico Nacional de México.", including the National Laboratory SEDEAM by sponsored electronics manufacturing.}%
\thanks{Gerardo~Marx~Chávez-Campos is with the Tecnológico Nacional de México - Instituto Tecnológico de Morelia, Av. Tecnológico 1500, Col. Santiaguito, CP 58120, México, and is the corresponding author:}
\thanks{gmarx\_cc@itmorelia.edu.mx}}

\begin{document}
\maketitle
\begin{abstract}
Weather is one of the main problems in implementing forecasts for photovoltaic panel systems. Since it is the main generator of disturbances and interruptions in electrical energy. It is necessary to choose a reliable forecasting model for better energy use. A measurement prototype was constructed in this work, which collects in-situ voltage and current measurements and the environmental factors of radiation, temperature, and humidity. Subsequently, a correlation analysis of the variables and the implementation of artificial neural networks were performed to perform the system forecast. The best estimate was the one made with 3 variables (lighting, temperature, and humidity), obtaining an error of 0.255. These results show that it is possible to make a good estimate for a photovoltaic panel system.
\end{abstract}


\section{Introduction}
Nowadays, energy consumption increases each year significantly. As a result, pollution increases by the use of fossil fuels during the production of complementary energy by energy companies \cite{el2016modeling}. Companies have incorporated alternative energy sources to reduce the impact and accomplish energy demand \cite{ener2019}. One of the most significant sources is solar energy, which has become the most popular alternative in the world \cite{powell2017hybrid}. Solar energy has been used to provide electricity for many years \cite{jayakumar2009resource}, by using Photo-Voltaic (PV) panels. The amount of current and power generated by a PV cell depends on external factors such as the environment and internal factors typical of the photo-voltaic system.

Specifically, the weather creates disturbances and interruptions in PV cells' electrical power \cite{muljadi2013pscad, tan2018overview, cui2017study, niemi2017analysis}. Due to this variability, it is necessary to implement reliable forecasts in the implementation of these PV systems to avoid penalties resulting from the differences between the programmed and produced energy  \cite{antonanzas2016review}.  Different forecasting models can be chosen based on the parameters to be analyzed, depending on particular needs \cite{mellit2008artificial}.

The use of artificial intelligence (AI) has increased considerably in recent years, due to its ability to model and solve complicated computational tasks\cite{mohammed2020multimodal}. In fact, AI algorithms in data collection systems help to improve the profitability of the measurement equipment \cite{alippi1998artificial} \cite{fogel2006machine}. Forecasting models are also good indicators for detecting the right moment to perform maintenance on the photovoltaic system and the distribution of the PV system. There are different methods of forecasting  PV energy. The very short-term method(from a few seconds to a few minutes) is used for the control and management of PV systems in the electricity market over micro-networks. The short-term method(48–72 hrs) for control of the energy system's operations, economic dispatch, and unit commitment, among others. The Medium-term (a few days to a week) and long-term methods (a few months to a year or more) are used to plan PV systems \cite{wan2015photovoltaic}.

Hence, several models and methods have been implemented to estimate the generated energy of a PV-systems. The advanced methods include diverse artificial intelligence and machine learning techniques, such as artificial neural networks (ANN), nearest neighbor-k (kNN), extreme learning machine (ELM), and support vector machine (SVM), to mention the most used \cite{russell2016artificial}. ML algorithms can be classified into three main groups:
\begin{enumerate}[(1)]
\item Supervised learning, where the algorithm creates relationships between input and output characteristics.
\item Unsupervised learning, in which the algorithm looks for patterns and rules to describe in a better way the data.
\item Reinforcement learning, which is used mainly with extensive data and reduces them for visualization or analysis purposes \cite{alpaydin2016machine}.
\end{enumerate}

The forecasting of power in PV systems is mainly based on ANN due to the complexity of the parameters involved. ANN are techniques that seek to emulate the human brain's behavior and generate responses for decision making \cite{bishop1995neural}. The fundamental part of each ANN is its element processor, the neuron. The neural networks gather these element processors with different methods, to respond to their different numerical needs \cite{hassoun1995fundamentals}.

Recently reported forecasting models are based on the ANNs. One type of model uses the ANN to estimate solar radiation (in specific cities) based on past weather data (temperature, humidity, and rain probability) together with radiation, with the estimated radiation, the models try to estimate the PV's output energy \cite{monteiro2013short, laouafi2015one, brenna2017solar}. Other models use the ANN to estimate  the produced energy directly using different inputs like: (i) humidity \cite{sangrody2017weather}; (ii) temperature, humidity, and rainfall \cite{verma2016data}; (iii) solar radiation and temperature \cite{tao2014distributed}; (iv) solar power and weather data \cite{sanjari2017probabilistic}. Eventually, an improved model considers data correlation and ANNs to select the most critical inputs \cite{menon2017correcting}. 
However, most models are based on databases available online or meteorological measurements not precisely in the same place as the photovoltaic system. Thus, the data information is insufficient from a zone or region, and in some cases, no continuous measurements are available. On the other hand, it is well known that semiconductors are extremely sensitive to high temperatures.

Therefore, the present paper proposes an IoT device to measure and log \textit{in-situ} data about the PV system. The solar radiation, solar panel's temperature-humidity, and the panel's electrical power (voltage and current) have been collected during ---120 days--- with a sampling frequency of 5 min, collecting more than 32,200 measurements. Then, measurements were used to train diverse ANN topologies and compare them with a Multiple Linear Regressor model.  The best topology has an error level of 0.255326464, which presents a reliable data model.

\section{Materials and Methods}
\begin{figure}[ht!]
  \centering
  \includegraphics[width=\linewidth]{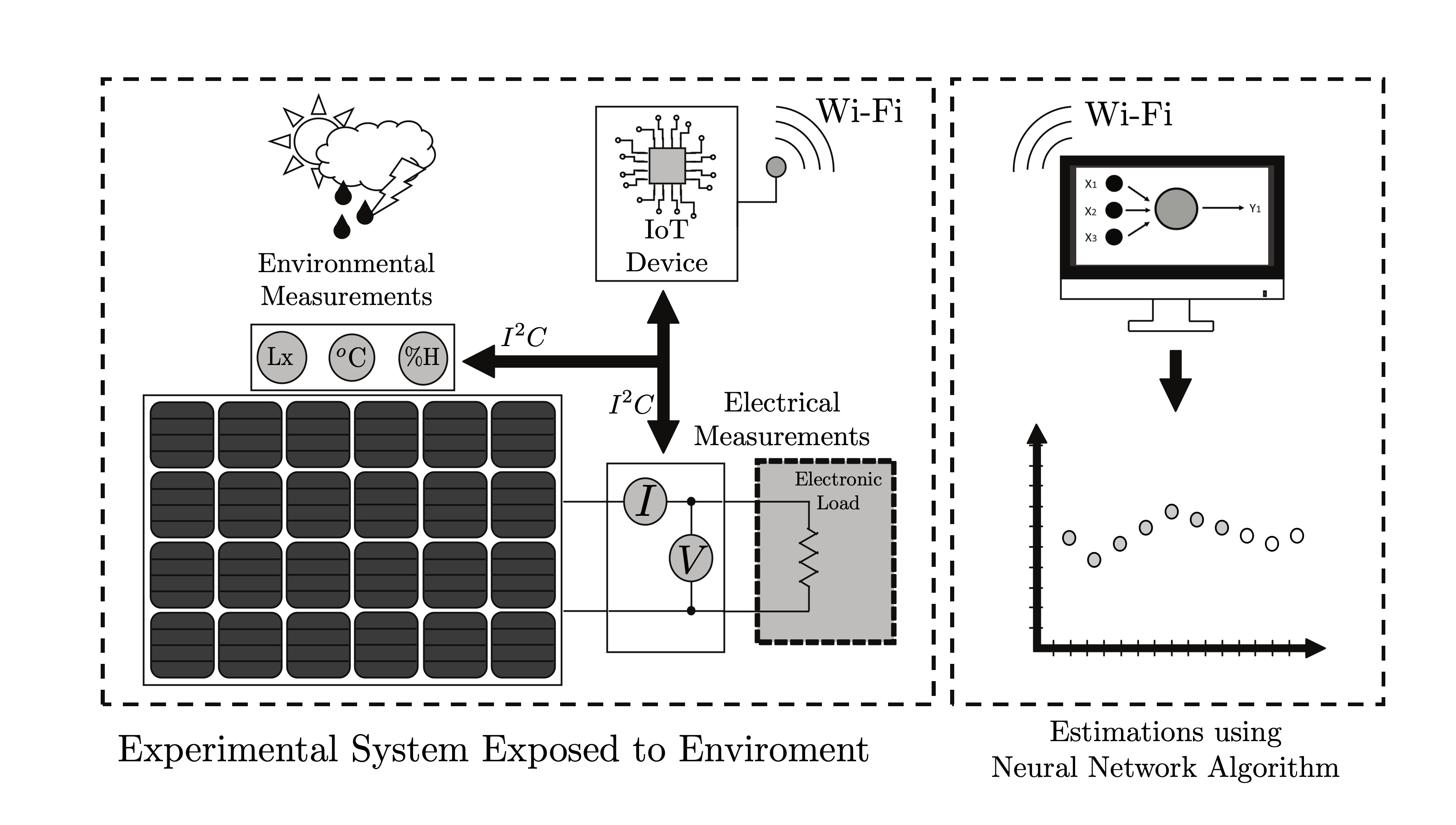} 
  \caption{Experimental setup for data gathering and analysis.}
  \label{fig:fig1marx}
\end{figure}

The present research methodology consists of acquiring and validating data from a photovoltaic system to perform the data analysis and compute electrical power forecasting values by artificial neural networks (see \autoref{fig:fig1marx}). Therefore, a measurement prototype has been designed to collect  \textit{in-situ} voltage and current measurements and environmental factors such as radiation levels, temperature, and humidity. Eventually, the system's behavior is obtained through the prototype's reading analysis, performing a validation process, and correlation of variables. Then, the regression model is obtained with a training set to finally implements the ANN.  The obtained estimations were evaluated against a test set using the Root-Mean Square Error (RMSE) as the primary measure, applying equation \ref{rmse}.

\begin{equation} \label{rmse}
RMSE = \sqrt{\frac{\sum \left ( x_{T}-x_{i} \right )}{N}}
\end{equation}

where $x_{T}$ is the estimated value, $x_{i}$ is the actual value and $N$ is the total number of measurements.

\subsection{Experimental setup}
\begin{figure}[h!]
  \centering
  \includegraphics[width=\linewidth]{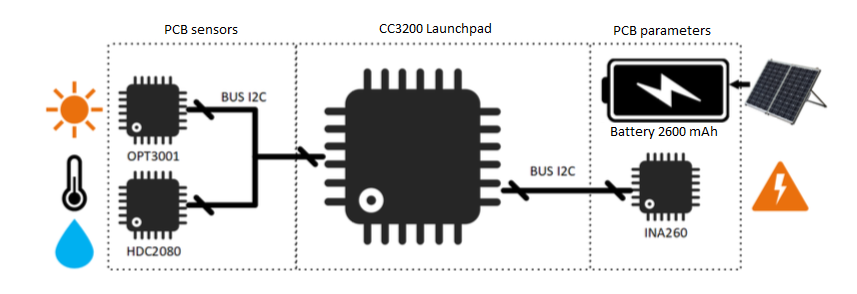}
  \caption{Experimental setup system for measurements.}
  \label{fig:experimentalSetup}
\end{figure}

Data has been collected using a stand-alone  IoT system embedded with three sensors with $I^2C$ communication. The IoT system also has embedded a web server to configure and manage the logged data \cite{jaisonitm}. The \autoref{fig:experimentalSetup} shows the experimental system conformed by three main sections. The left-must includes the environmental variable sensors, the OPT3001, and HDC2080 integrated circuits. The OPT3001 is a light sensor with a \SI{0.01}{\lux} resolution, including an upper limit of \SI{128}{\kilo\lux}. However,  an attenuating glass has been used to extend 55\% of the device limit; a calibration procedure was conducted using the commercial digital luxometer MASTECH ms6612. The HDC2080 sensor measures relative humidity (RH) and temperature. For temperature the sensor has a \SI{\pm2}{\degreeCelsius} resolution with ranges of \si{-40\degreeCelsius} to \si{85\degreeCelsius}, for RH the sensor gives measurements with resolution of $\pm 0.2\%$.
 
\subsection{Logged data analysis}
Previous to the forecasting process, logged data have been compared with similar commercial measurement systems to evaluate performance. According to the central limit theorem, our data is normal, so we performed the linear dependency measurement by the Pearson correlation method between measured variables to determine its importance during forecasting.

\begin{figure}[h!]
    \centering
    \includegraphics[width=\linewidth]{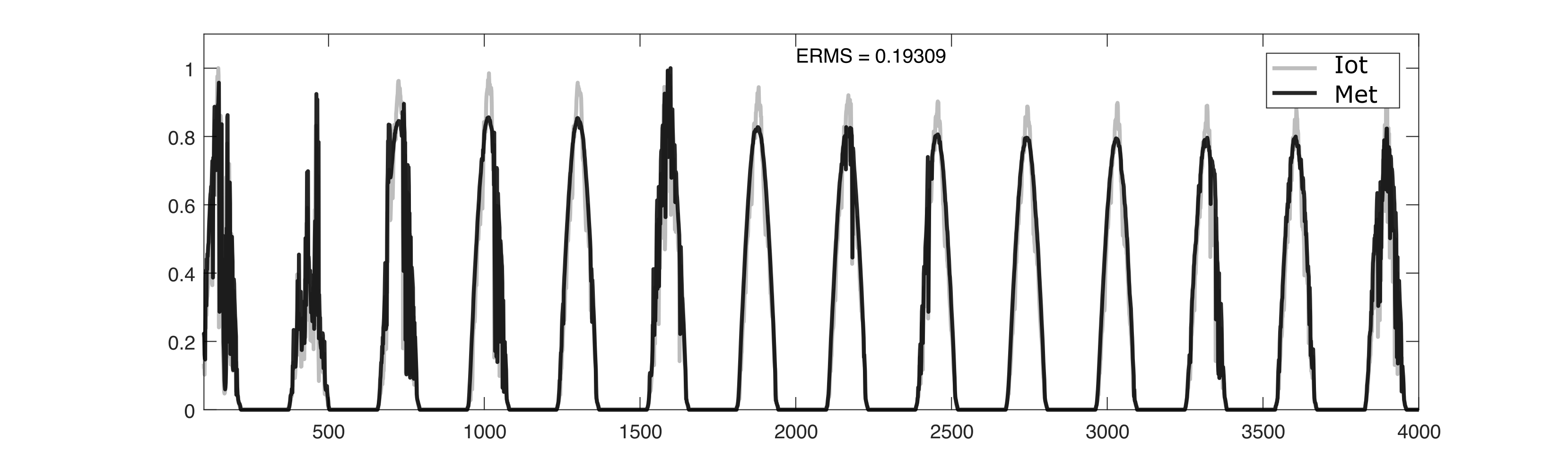}
    \caption{Data comparison between the meteorological station and the IoT system.}
    \label{fig:lux}
\end{figure}

The behavior comparison is made against data from the meteorological station of the Technological Institute of Morelia. Both data-sets are made up of solar radiation, humidity, and temperature, which correspond to the period 12th of November 2018 to the 6th of December 2018. Since measurements are originated by different equipment, different scale, and different locations (5 meters of difference), measurements were normalized using Eq. \ref{normal}:
\begin{equation} \label{normal}
x=\frac{x_{i}-x_{min}}{x_{max} - x_{min}}
\end{equation}
where $x_{i}$ is the actual value to be normalized, $x_{min}$ is the minimum value of the entire data-set, and $x_{max}$ is the maximum value. Fig. \ref{fig:lux} shows the measurements collected by the IoT system and the meteorological station only for the variable radiation. The small variations between data-sets can be attributed to the location of each system; however, the general performance is related.

\begin{figure*}[htb]
    \centering
    \includegraphics[width=\linewidth]{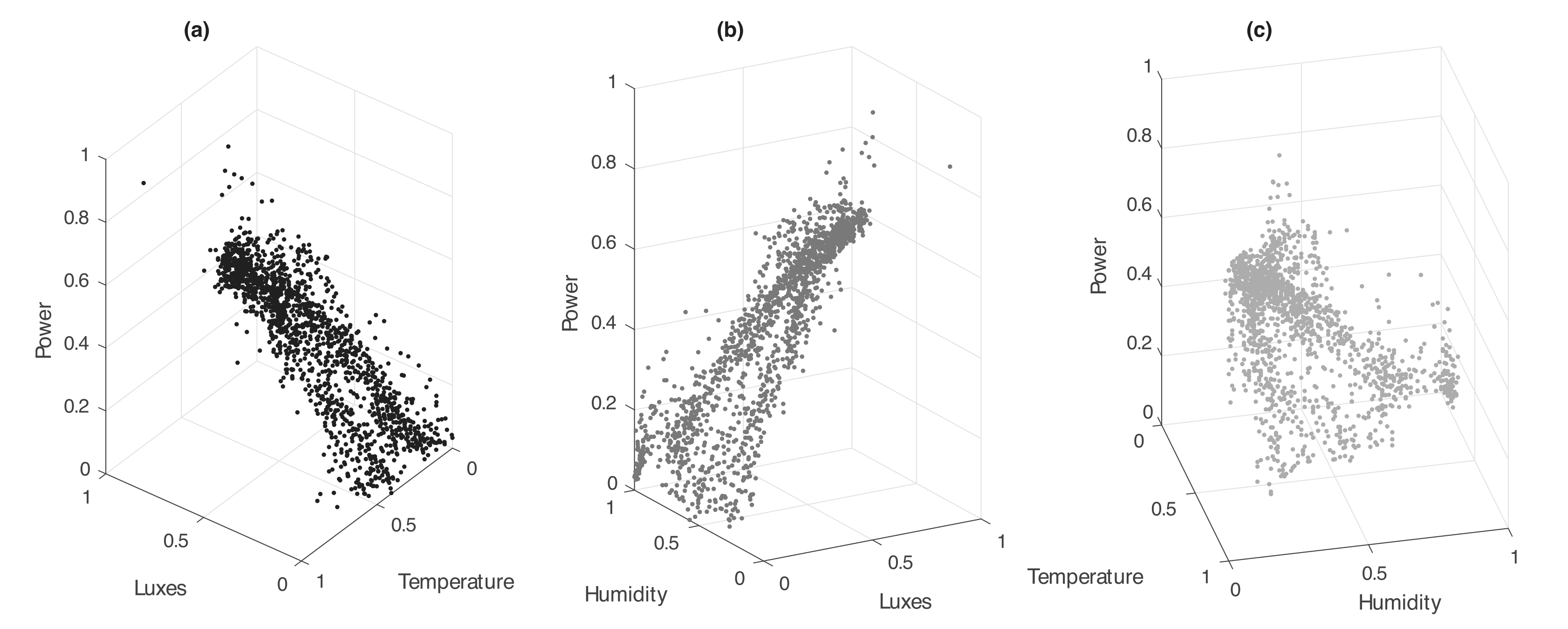}
    \caption{Correlation variables plots for (a) luxes, temperature and power; (b) humidity, luxes and power; and (c) temperature, humidity and power.}
    \label{fig:corr}
\end{figure*}

Fig. \ref{fig:corr} show three correlation plots: (a)luxes, temperature and power; (b)humidity, luxes and power; and (c)temperature, humidity and power. The three plots are made considering power variable as output. Therefore the third plot shows the lower correlation, meaning that humidity will have a lower impact on the final model.  

\section{Results}
The ANN performance analysis has considered cases with two and three input variables: (i) lighting, temperature, and humidity, (ii) lighting and temperature, (iii) lighting and humidity, and (iv) humidity and temperature; each option has been tested with different topologies. 

The training process was performed with random data from the 14th of January to the 10th of February 2019.  Table \ref{tab:compara} shows only the head-performing topologies based on the resulting RMSE obtained during the cross-correlation validation. On the table, the first column indicates the input variables. The second represents the topology used; the third and fourth ones are the maximum training cycles and error levels, respectively. The last column is the RMSE value. Notice that the maximum number of training cycles was defined based on the RMSE's performance during various training experiments. 

\begin{table*}[ht]
\caption{\textsc{Best ANN's topologies obtained for forecasting of PV systems.}}
\resizebox{\textwidth}{!}{
\begin{tabular}{ccccc}
\begin{tabular}[c]{@{}c@{}}\bfseries Input Variables\end{tabular} & \begin{tabular}[c]{@{}c@{}}\bfseries ANN\\ \bfseries Topology\end{tabular} & \multicolumn{1}{l}{\bfseries Training cycles} & \multicolumn{1}{l}{\bfseries Error level} & \multicolumn{1}{l}{\bfseries RMSE} \\
\toprule
lighting, temperature and humidity                        & 3:3:1                                                  & 5000                                & 0.1                             & 0.255                    \\
Lighting and Temperature                                  & 2:8:1                                                  & 5000                                & 0.1                             & 0.273                    \\
Lighting and Humidity                                     & 2:3:1                                                  & 5000                                & 0.1                             & 0.260                    \\
Temperature and humidity                                  & 2:7:1                                                  & 5000                                & 0.1                             & 1.52           \\         
\bottomrule
\end{tabular}}
\label{tab:compara}
\end{table*}

\begin{figure*}[h!]
    \centering
    \begin{subfigure}[b]{0.45\textwidth}
        \includegraphics[width=\linewidth]{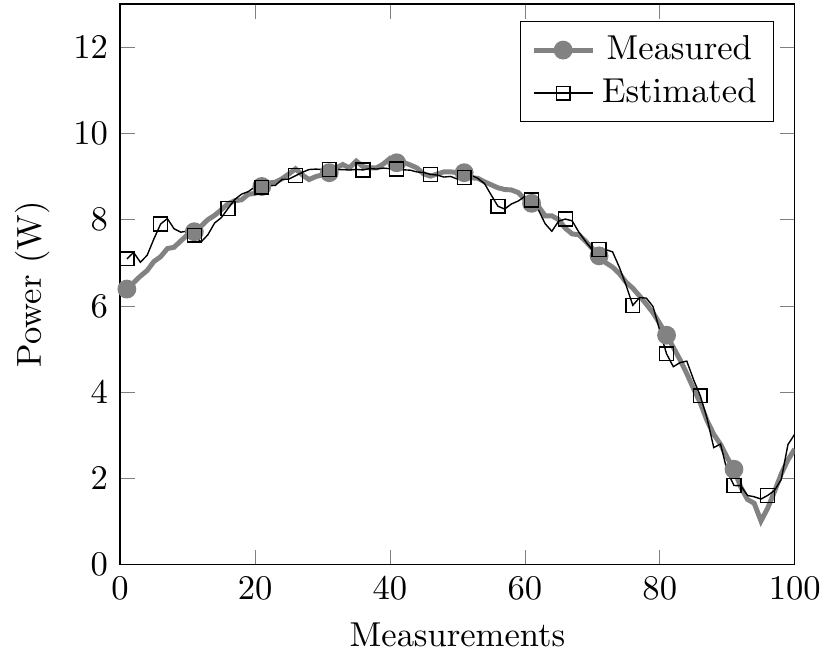}
        \caption{Estimation with 3 variables (Illumination, Temperature and Humidity) for a 3:3:1 topology}
        \label{fig:3:3:1}
    \end{subfigure}
    ~ 
    \begin{subfigure}[b]{0.45\textwidth}
        \includegraphics[width=\textwidth]{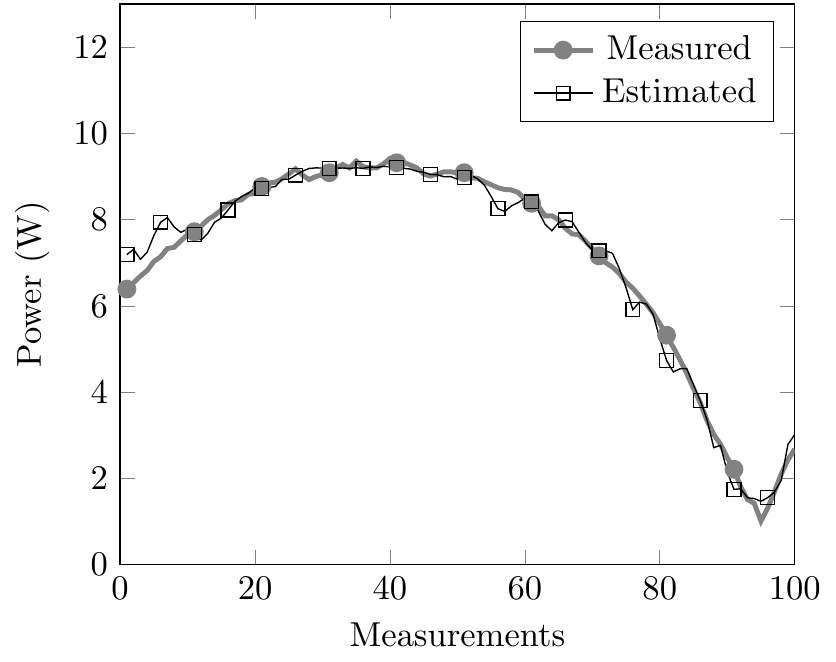}
        \caption{Estimation with 2 variables (lighting and temperature) for a 2:8:1 topology}
        \label{fig:2:8:1}
    \end{subfigure}
    \vspace{0.2in} 
    \begin{subfigure}[b]{0.45\textwidth}
        \includegraphics[width=\textwidth]{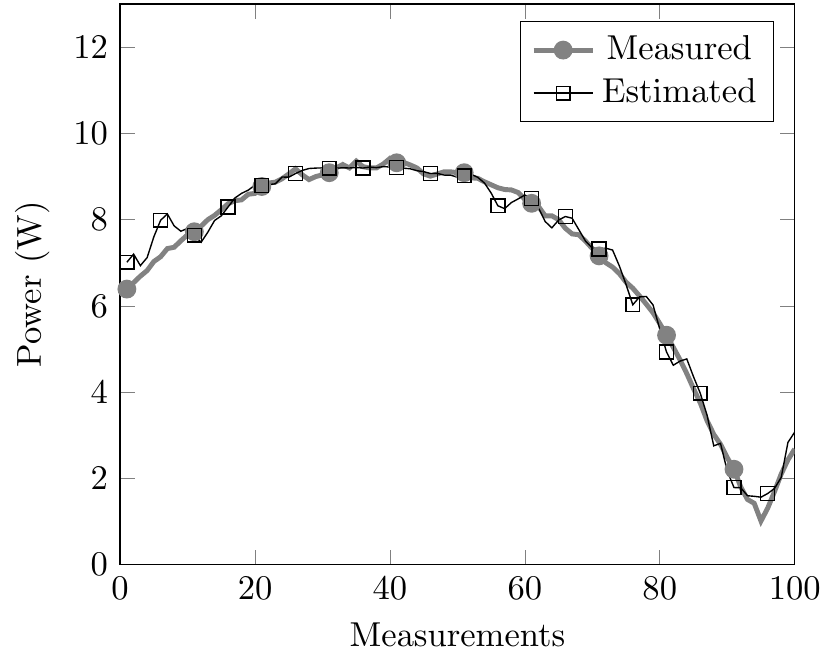}
        \caption{Estimation with 2 variables (Illumination and Humidity) for a 2: 3: 1 topology}
        \label{fig:2:3:1}
    \end{subfigure}
    ~ 
    \begin{subfigure}[b]{0.45\textwidth}
        \includegraphics[width=\textwidth]{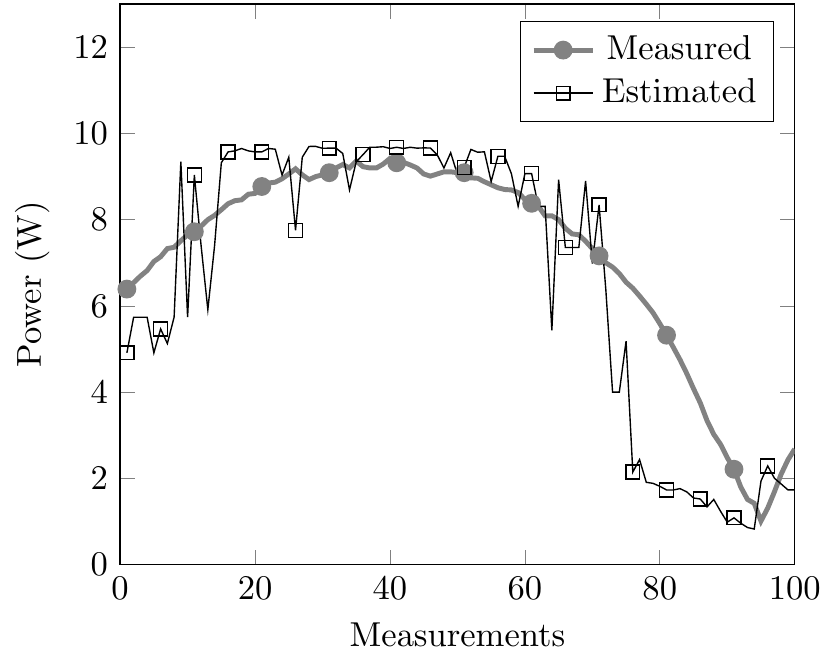}
        \caption{Estimation with 2 variables (Temperature and Humidity) for a 2:7:1 topology}
        \label{fig:2:7:1}
    \end{subfigure}
    \caption{Graphical comparison of estimated data against actual data for the best topology of two and three input variables}
    \label{fig:anns}
\end{figure*}

The first topology (Table \ref{tab:compara})  owns three computation elements in the input layer, three elements in the hidden layer, and one single output (3:3:1). This topology was evaluated with data that were randomly selected, and then the Easy-NN software was used to import the training set (700 measurements), validation (200), and test (100). Fig. \ref{fig:3:3:1} shows the comparison of estimated data with topology 3:3:1 vs. the test data-set. The computed level error for this topology was 0.255326464.  

Another good estimator has been implemented using only lighting and temperature, with a topology 2:8:1, resulting in an error level of 0.273086254; see Fig. \ref{fig:2:8:1}. In the case of the estimator based on the Illumination and Humidity, the best topology was 2:3:1, with an error level of 0.26061261; see Fig. \ref{fig:2:3:1}. Finally, the estimation with the variables Temperature and Humidity, the best performing topology was 2:7:1, with an error level of 1.522621379; see Fig. \ref{fig:2:7:1}. 

Notice that the best network applies the three variables. However, the optional networks can be used when some data is missing or corrupted. For this research, the humidity sensor was the most problematic due to the warmth variability, specifically in Morelia city, resulting in saturated measurements during the early morning periods. Nevertheless, when all variables are pleasant, the ANN can produce better estimations.

\begin{figure}[h!]
  \centering
  \includegraphics[width=\linewidth]{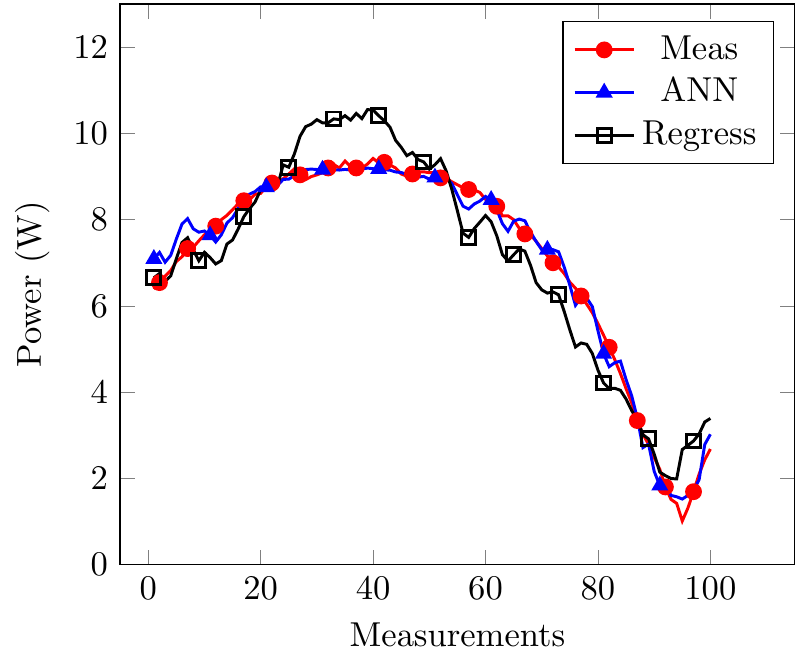}
  \caption{Comparative figure of the performances between the estimation by artificial neural network and regression model.}
  \label{fig:graf5_28}
\end{figure}

On the other hand, a Multiple Linear Regressor (MLR) has been developed with the same data-set with three variables. Figure \ref{fig:graf5_28} shows a comparison for one-day estimations of ANN, MLR, and real measurements. Notice that the ANN estimations are closely related to the photovoltaic system's real behavior, compared to the estimation made using the MLR model.

According to the central limit theorem, in large samples, the sampling distribution tends to be normal, regardless of the data \cite{ghasemi2012normality}. Therefore, an ANOVA analysis is carried out. The ANOVA of the MLR analysis was performed on lighting, temperature, and humidity to determine their importance, obtaining the Table \ref{tabla1}. The results of the ANOVA analysis gave a regression model with a confidence interval of 95 \%. The model has an approach of 91.77 \% of the real phenomenon.  This process is carried out to obtain learning with the right level of trust, which enables a suitable prediction of the power levels in the solar panels. 

\begin{table}[h!]
\caption{ANOVA Analysis of the $x$ variables (independent variables) and the $y$ variable (dependent variable).}
\begin{tabular}{cccccc}
\textbf{Source} & \textbf{DF} & \textbf{Adj SS} & \textbf{Adj MS} & \textbf{F-Value}            & \textbf{P-Value} \\ \hline
Regression      & 3           & 10159.50        & 3386.51         & \multicolumn{1}{l}{5895.20} & 0.000            \\
Illumination     & 1           & 2761.10         & 2761.10         & 4806.50                     & 0.000            \\
Temperature     & 1           & 6.80            & 6.78            & 11.81                       & 0.001            \\
Humidity         & 1           & 47.00           & 47.01           & 81.83                       & 0.000            \\
Error           & 1586        & 911.10          & 0.57            &                             &                  \\
Total           & 1589        & 11070.60        &                 &                             &                 
\end{tabular}
\label{tabla1}
\end{table}

\section{Discussion and Conclusions}

During this study, it was observed that the most suitable neural network topology changes according to the input variables because of the lowest RMSE value. Experiments were conducted with different training cycles, input information, number of neurons, and hidden layers to discern their execution, then choose the most appropriate for each set of data.  The prototype used has certain limitations, such as the resolution and ranges of the measurements and the flow of the readings. A comparison was made to verify the prototype's values besides a commercial station located at the Technological Institute of Morelia, Mexico. The RMSE calculation of 0.19309 determines that the prototype data is reliable for estimating.

The discrimination process, and creation of the data sets, were performed in Matlab\textregistered. The most suitable configuration was determined to carry out both instantaneous estimates and short-term forecasts. The most suitable topology for neural networks and their parameters (the number of computational elements, the number of hidden layers, the number of inputs and outputs, and training cycles) has been identified. The determination of each parameter starts from a proposed random topology, thus reaching the 3:3:1 configuration with 5,000 training cycles. This configuration allows knowing instantly how the solar panel will behave under normal conditions. The results obtained in this work show that the best option for the estimation is the 3:3:1 topology of the neural network, which uses three variables (lighting, temperature, and humidity), which allows estimating how much power you can get from a panel. As for forecasts, the best configuration is a 9:4:3:1 network. Even when using two variables gives a more significant error in the estimation than when using the three variables, this error being 0.30320 is reliable for the estimation.

This article introduces a solar forecasting algorithm based on the artificial neural network (ANN) model. The proposed model has a 3:3:1 topology with 5,000 training cycles. The clear sky model and meteorological data from the prototype are used to train the model. The prototype and the meteorological station of the Technological Institute of Morelia, located in Morelia, Mexico, were compared. 
The RMSE value confirms that it is possible to make sensible estimates using the lighting, temperature, and humidity data. Forthcoming work will direct on developing a more comprehensive multi-layer ANN model taking into account rainfall factors and time of day, as well as using a more massive data set to train the ANN model to achieve greater forecast accuracy. Also, the system's accuracy must be improved.

\section{Acknowledgments}
The authors would like to acknowledge the "Consejo Nacional de Ciencia y Tecnológia" (CONACYT) for the support received for developing this project  by supporting student 625015, also to The "Tecnológico Nacional de México" (TecNM) that supports the project 6127.17-P, and the "National Laboratory SEDEAM" by helping the development of the electronic prototypes. 

\section{Conflict of interest}
The authors declare that there is no conflict of interest.

\bibliographystyle{elsarticle-num}
\bibliography{bibliography.bib}

\begin{thebibliography}{10}
\expandafter\ifx\csname url\endcsname\relax
  \def\url#1{\texttt{#1}}\fi
\expandafter\ifx\csname urlprefix\endcsname\relax\def\urlprefix{URL }\fi
\expandafter\ifx\csname href\endcsname\relax
  \def\href#1#2{#2} \def\path#1{#1}\fi

\bibitem{el2016modeling}
I.~El~Kafazi, R.~Bannari, A.~Abouabdellah, Modeling and forecasting energy
  demand, in: Renewable and Sustainable Energy Conference (IRSEC), 2016
  International, IEEE, 2016, pp. 746--750.

\bibitem{ener2019}
Enerdata,
  \href{https://yearbook.enerdata.net/total-energy/world-consumption-statistics.html}{Global
  energy statistical yearbook} (2019).
\newline\urlprefix\url{https://yearbook.enerdata.net/total-energy/world-consumption-statistics.html}

\bibitem{powell2017hybrid}
K.~M. Powell, K.~Rashid, K.~Ellingwood, J.~Tuttle, B.~D. Iverson, Hybrid
  concentrated solar thermal power systems: A review, Renewable and Sustainable
  Energy Reviews 80 (2017) 215--237.

\bibitem{jayakumar2009resource}
P.~Jayakumar, Resource assessment handbook, Asia and Pacific Center for
  Transfer of Technology of the United Nations (2009).

\bibitem{muljadi2013pscad}
E.~Muljadi, M.~Singh, V.~Gevorgian, Pscad modules representing pv generator,
  Tech. rep., National Renewable Energy Laboratory (NREL), Golden, CO. (2013).

\bibitem{tan2018overview}
W.~C. Tan, L.~H. Saw, H.~San~Thiam, J.~Xuan, Z.~Cai, M.~C. Yew, Overview of
  porous media/metal foam application in fuel cells and solar power systems,
  Renewable and Sustainable Energy Reviews 96 (2018) 181--197.

\bibitem{cui2017study}
Y.~Cui, Y.~Su, Y.~Liu, Y.~Liu, D.~Smith, Study of variability metrics for solar
  irradiance and photovoltaic output, in: 2017 IEEE Power \& Energy Society
  General Meeting, IEEE, 2017, pp. 1--5.

\bibitem{niemi2017analysis}
A.~Niemi, M.~Lehtonen, H.~A.~R. AbdelHadi, Analysis of solar irradiance
  variations as a source of flicker associated with pv systems, in: 2017 IEEE
  PES Innovative Smart Grid Technologies Conference Europe (ISGT-Europe), IEEE,
  2017, pp. 1--6.

\bibitem{antonanzas2016review}
J.~Antonanzas, N.~Osorio, R.~Escobar, R.~Urraca, F.~J. Martinez-de Pison,
  F.~Antonanzas-Torres, Review of photovoltaic power forecasting, Solar Energy
  136 (2016) 78--111.

\bibitem{mellit2008artificial}
A.~Mellit, S.~A. Kalogirou, Artificial intelligence techniques for photovoltaic
  applications: A review, Progress in energy and combustion science 34~(5)
  (2008) 574--632.

\bibitem{mohammed2020multimodal}
S.~A. Mohammed, S.~Shirmohammadi, et~al., A multimodal deep learning-based
  distributed network latency measurement system, IEEE Transactions on
  Instrumentation and Measurement 69~(5) (2020) 2487--2494.

\bibitem{alippi1998artificial}
C.~Alippi, A.~Ferrero, V.~Piuri, Artificial intelligence for instruments and
  measurement applications, IEEE Instrumentation \& Measurement Magazine 1~(2)
  (1998) 9--17.

\bibitem{fogel2006machine}
D.~B. Fogel, Machine intelligence, IEEE instrumentation \& measurement magazine
  9~(3) (2006) 12--16.

\bibitem{wan2015photovoltaic}
C.~Wan, J.~Zhao, Y.~Song, Z.~Xu, J.~Lin, Z.~Hu, Photovoltaic and solar power
  forecasting for smart grid energy management, CSEE Journal of Power and
  Energy Systems 1~(4) (2015) 38--46.

\bibitem{russell2016artificial}
S.~J. Russell, P.~Norvig, Artificial intelligence: a modern approach, Malaysia;
  Pearson Education Limited,, 2016.

\bibitem{alpaydin2016machine}
E.~Alpaydin, Machine learning: the new AI, MIT press, 2016.

\bibitem{bishop1995neural}
C.~M. Bishop, et~al., Neural networks for pattern recognition, Oxford
  university press, 1995.

\bibitem{hassoun1995fundamentals}
M.~H. Hassoun, et~al., Fundamentals of artificial neural networks, MIT press,
  1995.

\bibitem{monteiro2013short}
C.~Monteiro, T.~Santos, L.~A. Fernandez-Jimenez, I.~J. Ramirez-Rosado, M.~S.
  Terreros-Olarte, Short-term power forecasting model for photovoltaic plants
  based on historical similarity, Energies 6~(5) (2013) 2624--2643.

\bibitem{laouafi2015one}
A.~Laouafi, M.~Mordjaoui, D.~Dib, One-hour ahead electric load and wind-solar
  power generation forecasting using artificial neural network, in: Renewable
  Energy Congress (IREC), 2015 6th International, IEEE, 2015, pp. 1--6.

\bibitem{brenna2017solar}
M.~Brenna, F.~Foiadelli, M.~Longo, D.~Zaninelli, Solar radiation and load power
  consumption forecasting using neural network, in: Clean Electrical Power
  (ICCEP), 2017 6th International Conference on, IEEE, 2017, pp. 726--731.

\bibitem{sangrody2017weather}
H.~Sangrody, M.~Sarailoo, N.~Zhou, N.~Tran, M.~Motalleb, E.~Foruzan, Weather
  forecasting error in solar energy forecasting, IET Renewable Power Generation
  (2017).

\bibitem{verma2016data}
T.~Verma, A.~Tiwana, C.~Reddy, V.~Arora, P.~Devanand, Data analysis to generate
  models based on neural network and regression for solar power generation
  forecasting, in: Intelligent Systems, Modelling and Simulation (ISMS), 2016
  7th International Conference on, IEEE, 2016, pp. 97--100.

\bibitem{tao2014distributed}
Y.~Tao, Y.~Chen, Distributed pv power forecasting using genetic algorithm based
  neural network approach, in: Advanced Mechatronic Systems (ICAMechS), 2014
  International Conference on, IEEE, 2014, pp. 557--560.

\bibitem{sanjari2017probabilistic}
M.~J. Sanjari, H.~Gooi, Probabilistic forecast of pv power generation based on
  higher order markov chain, IEEE Transactions on Power Systems 32~(4) (2017)
  2942--2952.

\bibitem{menon2017correcting}
V.~P. Menon, S.~Lokhande, Y.~K. Bichpuriya, Correcting forecast of utility's
  demand with increasing solar pv penetration, in: Innovative Smart Grid
  Technologies-Asia (ISGT-Asia), 2017 IEEE, IEEE, 2017, pp. 1--6.

\bibitem{jaisonitm}
J.~Corona-Ventura, Characterization of solar panels using the internet of
  things (IoT), Morelia Institute of Technology, 2019.

\bibitem{ghasemi2012normality}
A.~Ghasemi, S.~Zahediasl, Normality tests for statistical analysis: a guide for
  non-statisticians, International journal of endocrinology and metabolism
  10~(2) (2012) 486.

\end{thebibliography}

\end{document}